\documentclass[a4paper,11pt]{article}
\usepackage{jinstpub} 
\usepackage{lineno}

\title{Optimizing Neon-based Gas Mixtures for Two-stage Amplification Fast-timing Micromegas Detectors}

\author[a,b]{Y. Meng}
\author[a,b]{X. Wang}
\author[a,b,1]{J. Liu\note{Corresponding author.}}
\author[a,b]{M. Shao}
\author[a,b]{Z. Zhang}
\author[a,b]{Y. Zhou}
\affiliation[a]{State Key Laboratory of Particle Detection and Electronics, University of Science and Technology of China, Hefei 230026, China}
\affiliation[b]{Department of Modern Physics, University of Science and Technology of China, Hefei 230026, China}

\emailAdd{liujianb@ustc.edu.cn}

\abstract{Working gas components significantly impact the performance of gaseous detectors. A fast-timing Micromegas detector with two-stage amplification is prone to notable deterioration of uniformity when scaled up. This paper presents a simulation study based on Garfield++ that aims to enhance the performance of such detectors by exploring different gas mixtures. The properties of various gas compositions and their impact on detector performance including gain uniformity and time resolution were investigated in the simulation study. The gain uniformity and single-photon time resolution of the detector were evaluated in tests using a multi-channel PICOSEC Micromegas (MM) prototype with different gas mixtures. The experimental results are consistent with the findings of the simulation. Both simulation and experimental results indicate that a higher concentration of neon improves the detector’s gain uniformity, while the impact of gas mixtures on time resolution should also be considered as a critical performance indicator. The study presented in this paper offers valuable insights for improving uniformity in large-area PICOSEC MM detectors and optimizing overall performance.}


\begin{document}
\maketitle
\flushbottom

\section{Introduction}
\begin{figure}[htbp]
\centering
    \begin{minipage}{0.45\textwidth} 
        \centering 
        \includegraphics[width=\textwidth]{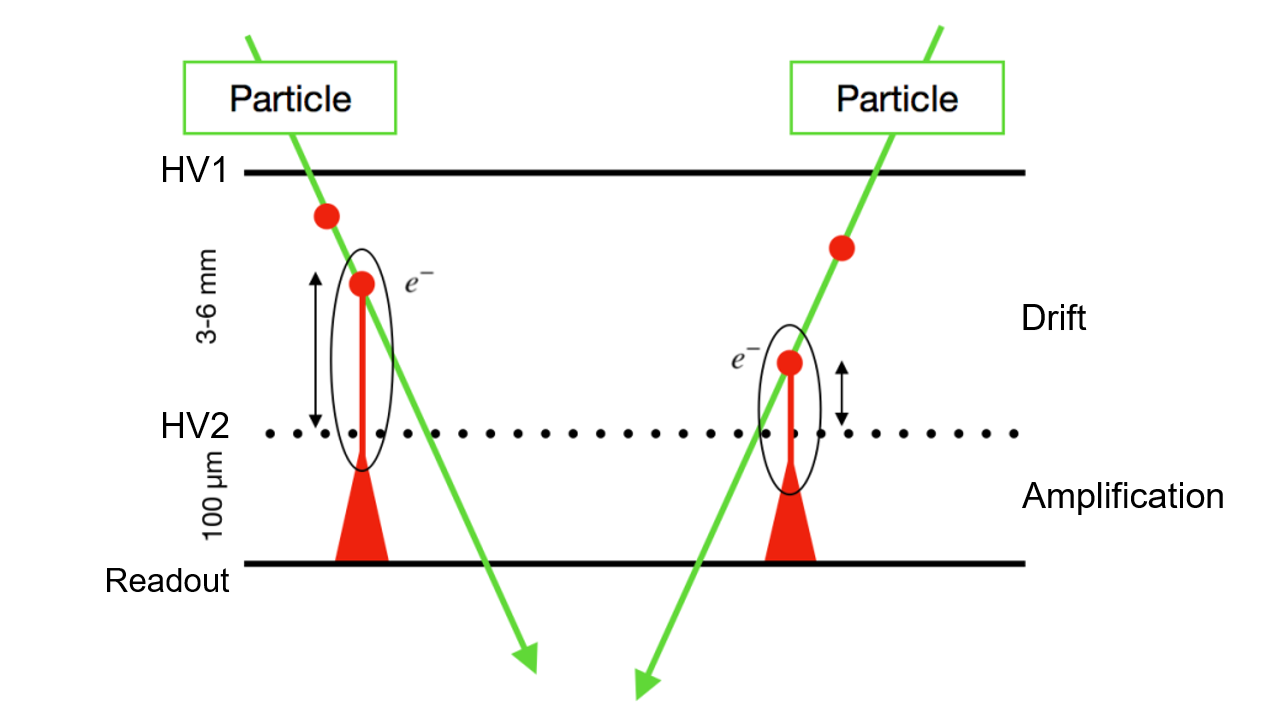} 
        \vspace{2mm} 
        \textbf{(a)} 
    \end{minipage}
    \qquad
    \begin{minipage}{0.45\textwidth}
        \centering 
        \includegraphics[width=\textwidth]{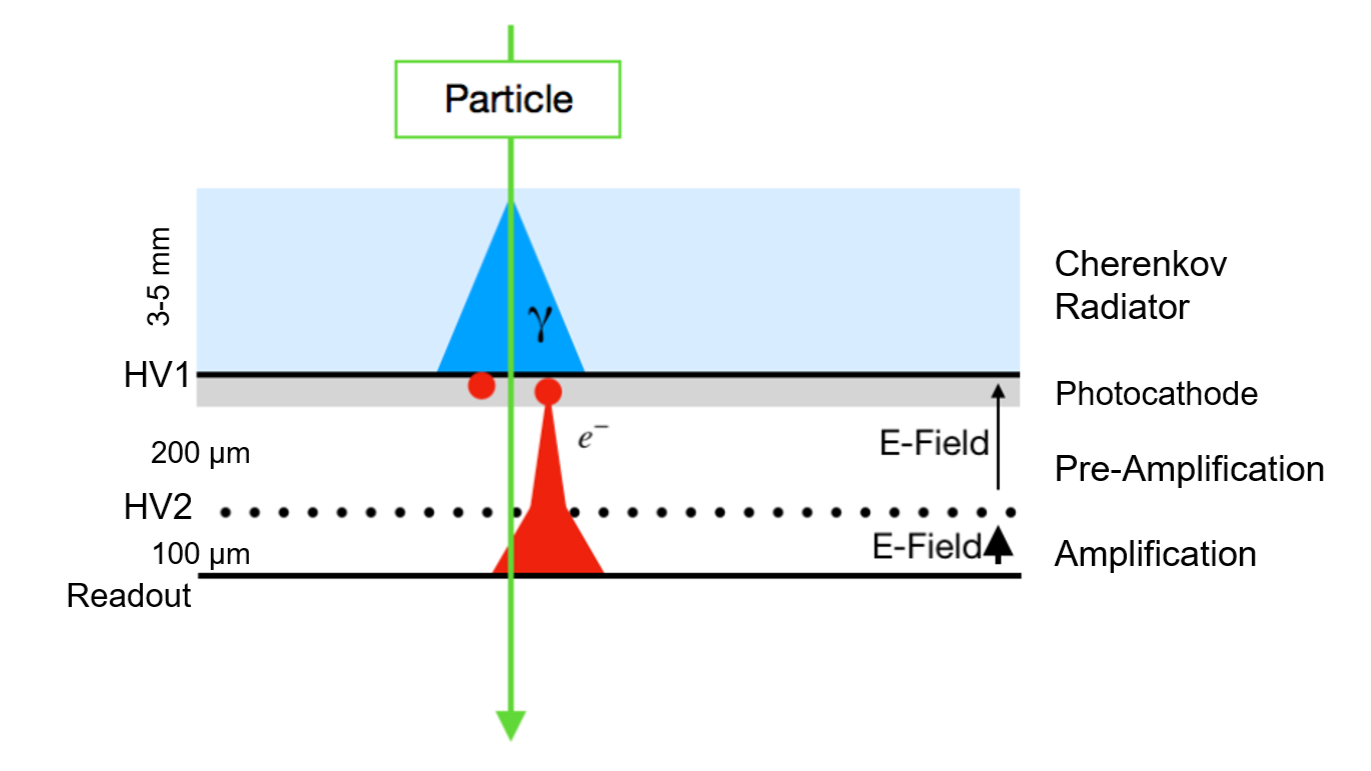} 
        \vspace{2mm} 
        \textbf{(b)} 
    \end{minipage}
\caption{(a) Scheme of Micromegas. (b) Scheme of PICOSEC MM.\label{fig:f1}}
\end{figure}

The MicroMeshGaseous Structure (Micromegas) \cite{a} is a micro-pattern gaseous detector that has gained widespread use in nuclear and particle physics experiments due to its notable advantages: excellent spatial and energy resolution, high-rate capabilities, and cost-effective operation over large areas. The classic Micromegas features two regions split by a thin mesh: a drift gap for primary ionization and an amplification gap with a high electric field (50-100kV/cm) \cite{d} to multiply ionized electrons, as shown in figure \ref{fig:f1} (a).
Operating at gains higher than $10^4$, induced signals are generated by both ionized electrons and ions. Extensive research has been conducted to maintain uniformity over large areas, leading to various fabrication techniques, such as bulk MM \cite{c} and thermal bonding MM \cite{d}. An overall uniformity of 10\% can be achieved among the detector channels by strictly controlling the production procedure \cite{e,f}.

\begin{figure}[htbp]
\centering
\includegraphics[width=.6\textwidth]{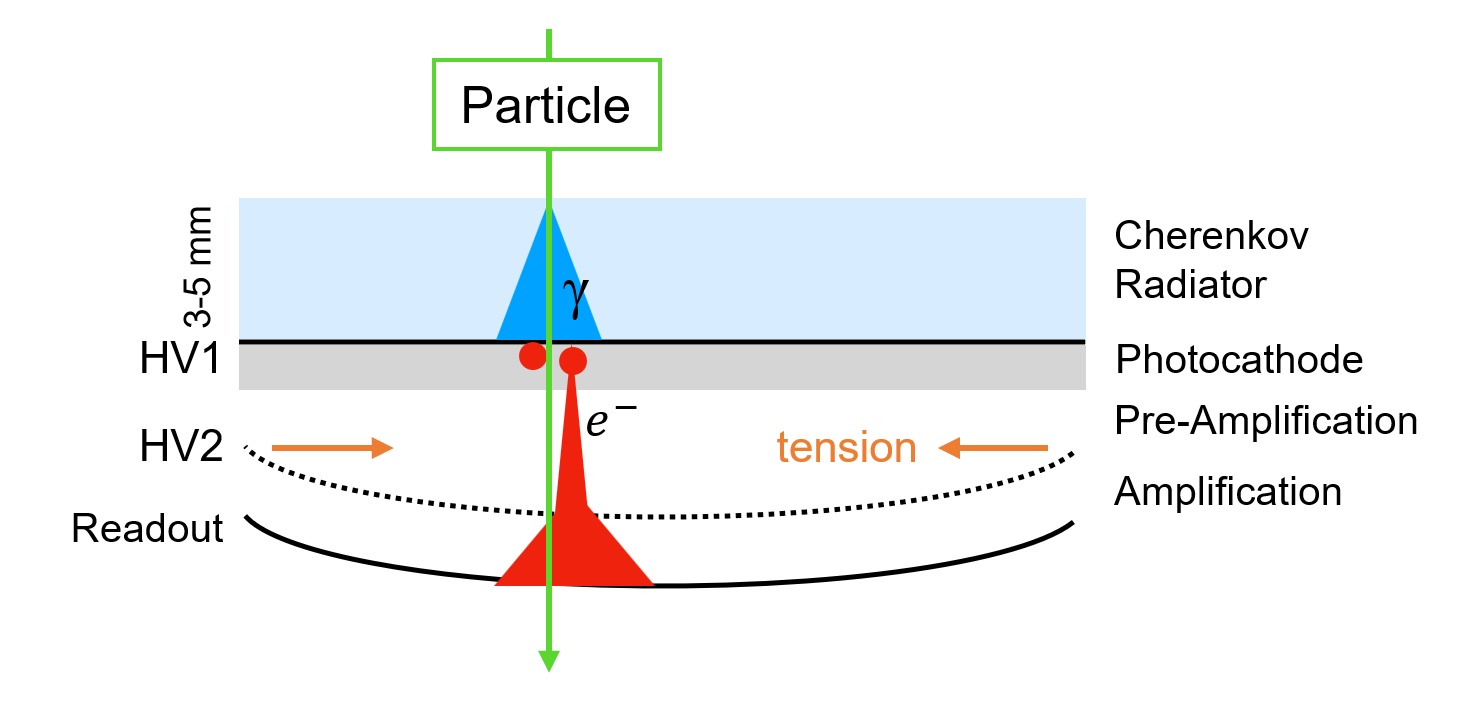}
\caption{PICOSEC MM structure after deformation caused by mesh tension.\label{fig:f13}}
\end{figure}

A two-stage amplification Micromegas, known as the PICOSEC Micromegas \cite{g}, was developed based on the structure of classic Micromegas in order to achieve superior timing performance. In this design, a strong electric field of the same magnitude as that in the amplification gap is applied to the drift gap, allowing it to perform a pre-amplification function, also known as the pre-amplification (pre-amp) gap. The  PICOSEC Micromegas couples a Cherenkov radiator and a semi-transparent photocathode with a two-stage amplification Micromegas, as shown in figure \ref{fig:f1} (b) \cite{AUNE2021165076}. When particles arrive, photoelectrons are produced simultaneously on the photocathode, and the primary ionization in the working gas is limited within the very thin drift gap, such that the time jitter due to primary ionization in this structure is highly eliminated. However, the two-stage amplification micromegas faced significant uniformity deterioration while scaling up to large areas. Despite their excellent time resolution performance, notable non-uniformity was observed, arising from the detector’s inherent structural property and production constraints \cite{h}. Previous studies indicated that this non-uniformity mainly stems from the deformation of the PCB caused by tension exerted from the mesh during the fabrication proccess \cite{i}. The schematic diagram of the PICOSEC MM after deformation due to mesh stretching is shown in figure \ref{fig:f13}, resulting in an uneven pre-amp gap. Such a phenomenon can also occur in a standard Micromegas detector; however, it has minimal impact on gain uniformity because the multiplication process is confined to the amplification region. Even if the PCB is distorted by the mesh, the mesh and the PCB bend and deform together, maintaining a relatively uniform amplification gap thickness. This is not the case for a two-stage amplification Micromegas, where multiplication also occurs in the pre-amp gap. In this scenario, a thinner pre-amp gap exacerbates the effects of tension-induced deformation. Even slight deformation can cause the detector gain to be non-uniform, exhibiting a trend of being higher at the edges and lower in the center. Efforts have been made to enhance the mechanical strength of the PCB, such as introducing ceramic into the board \cite{utrobicic2022large}, yet completely eliminating the mechanical deformation of the PCB remains challenging. Such effects of uniformity could potentially impact the detector’s time response, particularly as the detection area further expands in future developments.

The selection of working gas can influence the performance of MPGDs by affecting processes such as gas molecule ionization and electron transfer within the detector. Optimizing gas compositions can be a valuable approach to minimizing the non-uniformity of the detector. The original working gas in PICOSEC MM (80\% Ne, 10\% $\text{CF}_4$  and 10\% $\text{C}_2\text{H}_6$) has demonstrated superior timing performance in previous studies \cite{j}. However, the proportion has not been optimized for the interest of achieving good uniformity in the detector. Thus, a comprehensive investigation into the impact of various gas mixtures is essential for better understanding the detector’s characteristics and further optimizing its performance.

In this paper, a simulation based on Garfield++ \cite{k} was developed to study the influence of several gas compositions on the detector’s uniformity and time resolution. Meanwhile, experiments were taken with a 10$\times$10 $\text{cm}^2$ multi-channel PICOSEC MM under different gas mixtures to verify the simulation result. 

\section{Simulation and Experimental Setup}
\subsection{Simulation Setup}
\begin{figure}[htbp]
\centering
    \begin{minipage}[t]{0.45\textwidth} 
        \centering 
        \includegraphics[width=\textwidth]{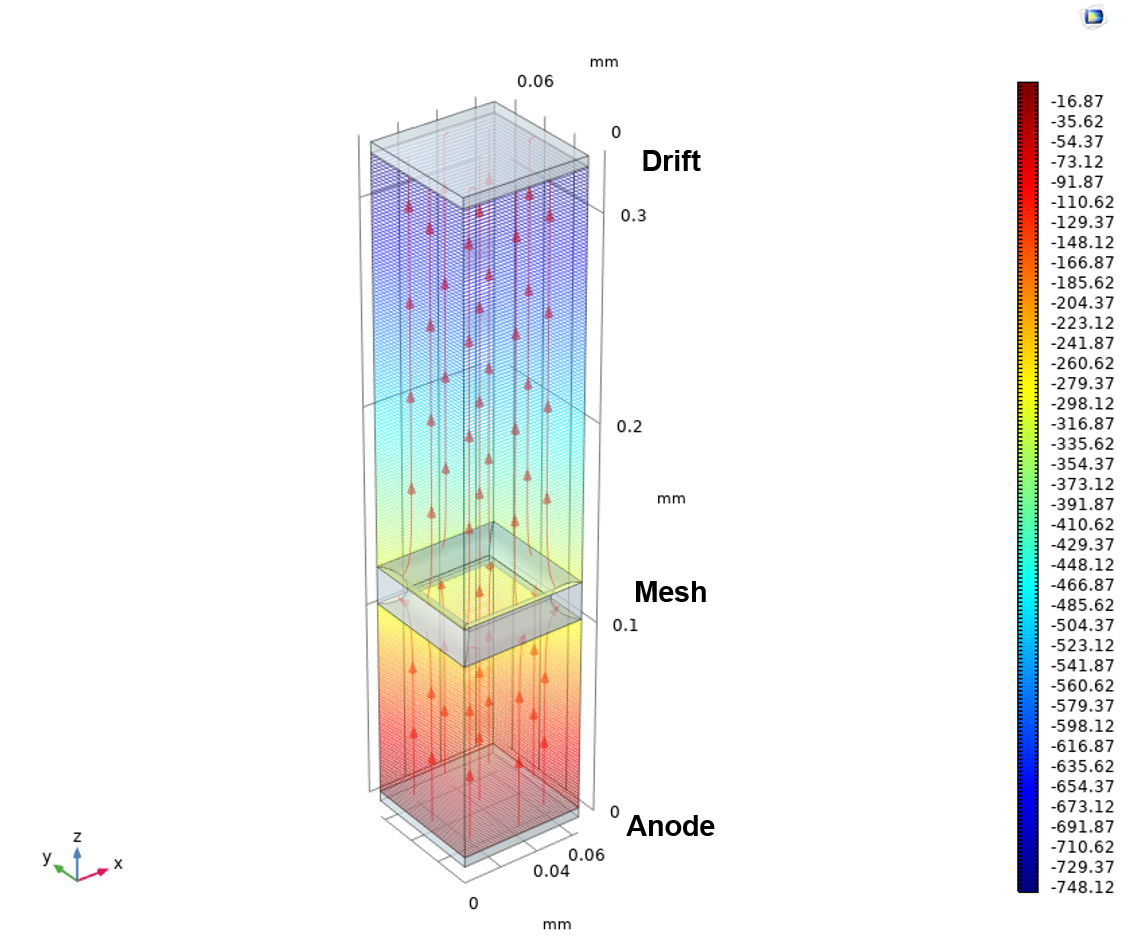} 
        \vspace{2mm} 
        \textbf{(a)} 
    \end{minipage}
    \qquad
    \begin{minipage}[t]{0.35\textwidth}
        \centering 
        \includegraphics[width=\textwidth]{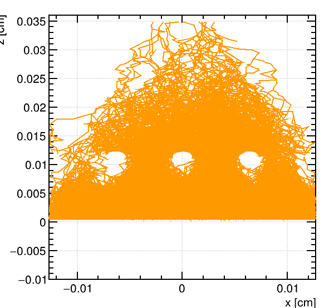} 
        \vspace{2mm} 
        \textbf{(b)} 
    \end{minipage}
\caption{(a) Simulation model built in COMSOL and the electric field generated. (b) Electron avalanche and transport processes in Garfield++.\label{fig:f2}}
\end{figure}

A simulation was established to study the influence of several gas mixtures on the detector’s uniformity using the Monte Carlo method. The PICOSEC MM structure in figure \ref{fig:f1} (b) is used to build the model in the simulation. The amplification gap thickness is fixed at 100 \textmu m, while the pre-amp gap thickness varies with a reference of 170 \textmu m to mimic actual variation. The mesh type established here is consistent with the actual implementation, featuring a density of 400 LPI and a wire diameter of 19 \textmu m. Finite element method (FEM) programs are used to build the amplification structure and calculate the electric field map. Figure \ref{fig:f2} (a) depicts the amplification structure model built in COMSOL, with contour lines showing the potential distribution and arrows indicating the electric field strength. Garfield++ is employed to simulate charge avalanche and transport processes, as well as to calculate the induced signal from the photoelectrons. To simulate the primary photoelectrons emitted at the photocathode, electrons with a slight momentum of 0.1 eV in a random direction are placed at the drift plane. Figure \ref{fig:f2} (b) shows the electron avalanche and transport processes performed in Garfield++, where the primary electrons are pre-amplified and pass through the mesh for further multiplication. Penning transfer, which influences total ionization, is also taken into account and set to 0.7 for all gas mixtures. The average gain is derived by filling the avalanche size of each event into a histogram and fitting it with a Polya distribution \cite{sohl2020development, schindler2012microscopic}, which is then used to estimate the detector’s uniformity. The gas mixtures chosen are based on the standard gas mixture of 80\% Ne, 10\% $\text{CF}_4$  and 10\% $\text{C}_2\text{H}_6$, varying the composition of Ne from 80\% to 96\% while maintaining a 1:1 ratio of the remaining $\text{CF}_4$ and $\text{C}_2\text{H}_6$. The Townsend coefficient for each mixture was then calculated using the libraries in Magboltz.

The detector’s time resolution was likewise simulated since it is regarded as a significant indicator when estimating the detector's performance. To obtain the time resolution, information on the signals is then extracted and analyzed. The simulation employed a process identical to the experimental tests to derive time resolution. The raw signals extracted in the simulation were used to mimic the response of the signals coming after a fast-current amplifier used in the test. The reference time is determined as the moment when the initial electron is generated. Signal arrival time (SAT) is derived by implementing 20\% constant fraction discrimination (CFD) on the signal's rising edge. Subsequently, time resolution is determined by fitting the SAT distribution with a Gaussian function. More details of the signal analysis method can be found in \cite{g}.

\subsection{Experimental Setup}
\begin{figure}[htbp]
\centering
    \begin{minipage}[t]{0.55\textwidth} 
        \centering 
        \includegraphics[width=\textwidth]{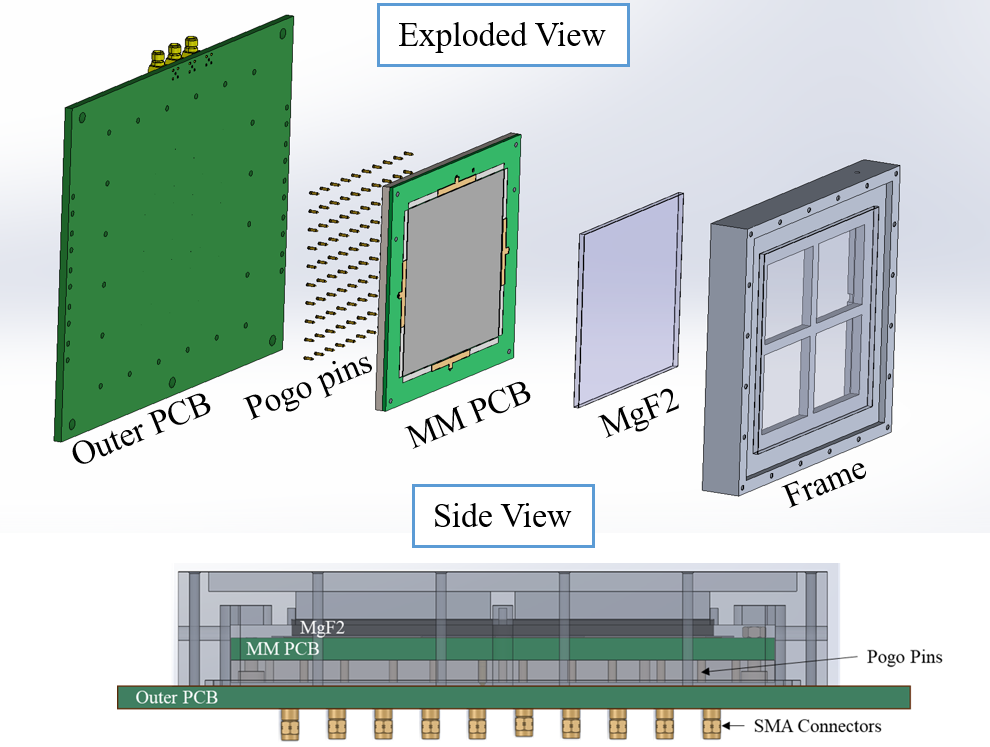} 
        \vspace{2mm} 
        \textbf{(a)} 
    \end{minipage}
    \qquad
    \begin{minipage}[t]{0.35\textwidth}
        \centering 
        \includegraphics[width=\textwidth]{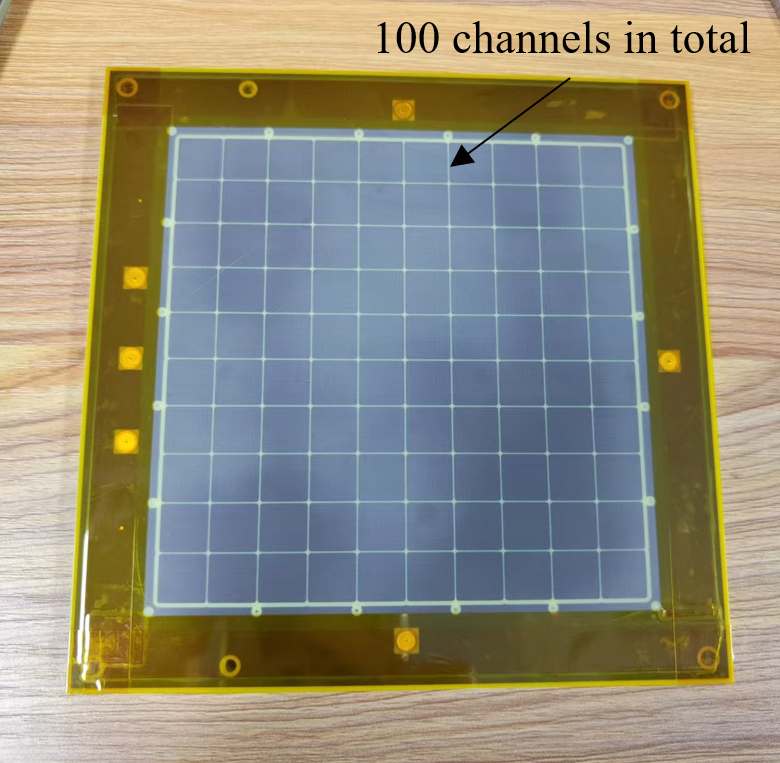} 
        \vspace{3mm} 
        \textbf{(b)} 
    \end{minipage}
\caption{(a) Scheme of the 10$\times$10 $\text{cm}^2$ multi-channel PICOSEC MM layout. (b) MM PCB before adhesion of the mesh.\label{fig:f3}}
\end{figure}

The experimental test utilized a 100-channel PICOSEC MM with 10$\times$10 $\text{cm}^2$  effective area. The detector’s main structure consists of a gas frame, a 104$\times$104 $\text{mm}^2$ $\text{MgF}_2$ as Cherenkov Radiator, an MM PCB to form Micromegas and an outer PCB to exert signals, as shown in figure \ref{fig:f3} (a). Diamond-like carbon (DLC) with a thickness of 3 nm is coated on the $\text{MgF}_2$ as the photocathode \cite{wang2024novel}. Figure \ref{fig:f3} (b) shows the 100 channels arranged in a 10$\times$10 grid on the MM PCB. The Micromegas structure in the detector is formed by coating germanium on the MM PCB as a resistive electrode and applying the thermal bonding method to stretch the stainless-steel mesh on top \cite{m,n}. The thickness of the pillars under the mesh is to form the amplification gap. The spacer is placed on top of the mesh between MM PCB and $\text{MgF}_2$ to define the pre-amp gap. The thickness of each gap is consistent with the parameters used in the simulation. Over one hundred pogo pins are placed to extract signals generated from the MM PCB to the Outer PCB, which can eventually be read from the SMA connectors outside the Outer PCB.

The uniformity of the 100-channel PICOSEC MM was tested using different gas mixtures with varying neon concentrations. Single photons emitted from a candle are employed to assess detector gain across multiple channels. Gas mixtures are blended using three flowmeters, each monitoring the percentages of Ne, $\text{CF}_4$  and $\text{C}_2\text{H}_6$. The signals from PICOSEC MM are directed through a charge sensitive pre-amplifier (ORTEC 142AH), a spectroscopy amplifier (ORTEC 671) and recorded in a multichannel analyzer (MCA8000D).

\begin{figure}[htbp]
\centering
\includegraphics[width=.8\textwidth]{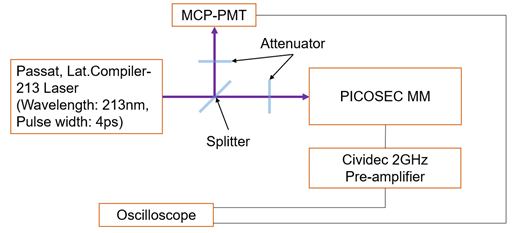}
\caption{Schematic diagram of the laser test setup.\label{fig:f4}}
\end{figure}

Time resolution for the single photoelectron (SPE) of the detector was also evaluated in the laboratory using a UV light laser. The laser test setup is illustrated in figure \ref{fig:f4}. A laser (COMPILER-213) manufactured by Passat Ltd generates a pulse laser beam with a width of 4 ps and a wavelength of 213 nm. The UV light emitted from the laser is split into two beamlines, and their intensities are tuned using two variable optical attenuators. One beam line is directed towards a Hamamatsu MCP-PMT (model R3809U-50) \cite{o}, serving as a time reference detector with a time resolution of better than 10 ps, as well as a trigger detector. The other beamline is attenuated to a single photon level to generate signals in the detector. The SPE signals generated from the detector are amplified by a current amplifier (CIVIDEC 2 GHz), then recorded on an oscilloscope (LeCroy, 4 GHz bandwidth, 20 Gsamples/s), along with the signal from the MCP-PMT. The time resolution of the detector is determined by fitting the distribution of the time difference between the MCP-PMT and the measured detector using the same data analysis method employed in the simulation.

\section{Gain Uniformity}
\subsection{Simulation of Gain}
\begin{figure}[htbp]
\centering
\includegraphics[width=.8\textwidth]{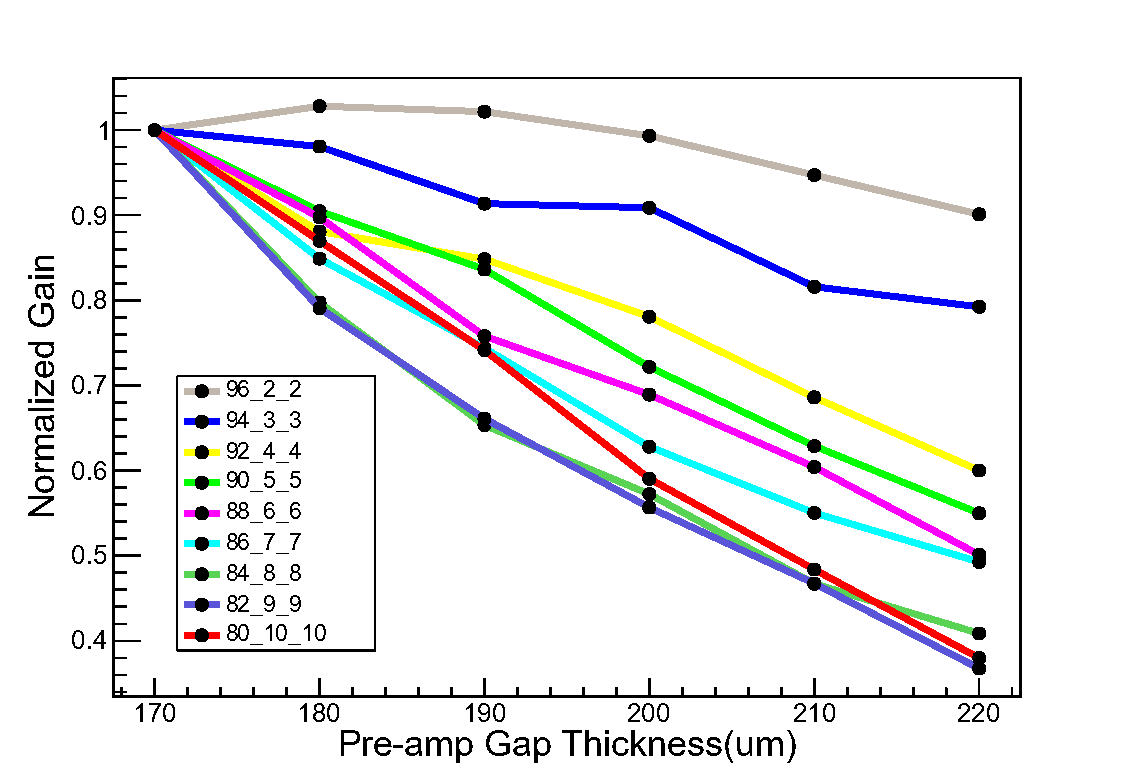}
\caption{Simulation of normalized gain versus pre-amp gap thickness across various gas mixtures.\label{fig:f5}}
\end{figure}

Detector gain with various pre-amp gap thicknesses was simulated across different gas mixtures since the non-uniformity in gain arises from variation in pre-amp gap thickness. In the simulation, the thickness of the pre-amp gap varies from 170 \textmu m to 220 \textmu m, with the high voltage applied on the pre-amp gap and amplification gap set at 350 V and 300 V, respectively. Figure \ref{fig:f5} illustrates the simulated gas gain as a function of pre-amp gap thickness across various gas mixtures, with the highest gain within a mixture set to 1. While the gain increases within the richer neon concentration, the variation curve becomes flatter as the pre-amp gap thickness changes. Notably, the gas mixture containing 96\% neon exhibits superior performance, with normalized gain remaining above 0.9 times despite pre-amp gap variations, compared to the decline to 0.4 times seen in the 80\% neon mixture. It should be noted that, in real conditions, the Penning transfer rate varies slightly depending on the gas mixture, but this variation has not been considered here. This simulation indicates that the detector gain is less sensitive to the non-flatness of the pre-amp gap with higher neon concentration. In this case, the selection of gas mixtures rich in neon is justified for enhancing gain uniformity.

The phenomena observed in simulation results can be explained using the Townsend avalanche theory, which describes avalanche processes in gas mediums. A further simulation was conducted in Garfield++ using the libraries in Magboltz to compute the transport properties for various gas mixtures. The number of ionizing collisions caused by an electron per unit distance as it moves through a gas under the influence of an electric field is referred to as the First Townsend Coefficient ($ \alpha $). Penning transfer occurs when excited neon atoms collide with surrounding molecules, transferring their excitation energy and ionizing those molecules, thereby enhancing the overall ionization in a gas mixture. The Effective Townsend Coefficient ($ \alpha _{eff} $) takes into account both direct ionization and additional ionization caused by the Penning transfer. It can be expressed as $ \alpha _{eff} = \alpha + \beta $, where $ \beta $ represents the additional ionization contribution due to the Penning Transfer.

\begin{figure}[htbp]
\centering
\includegraphics[width=.8\textwidth]{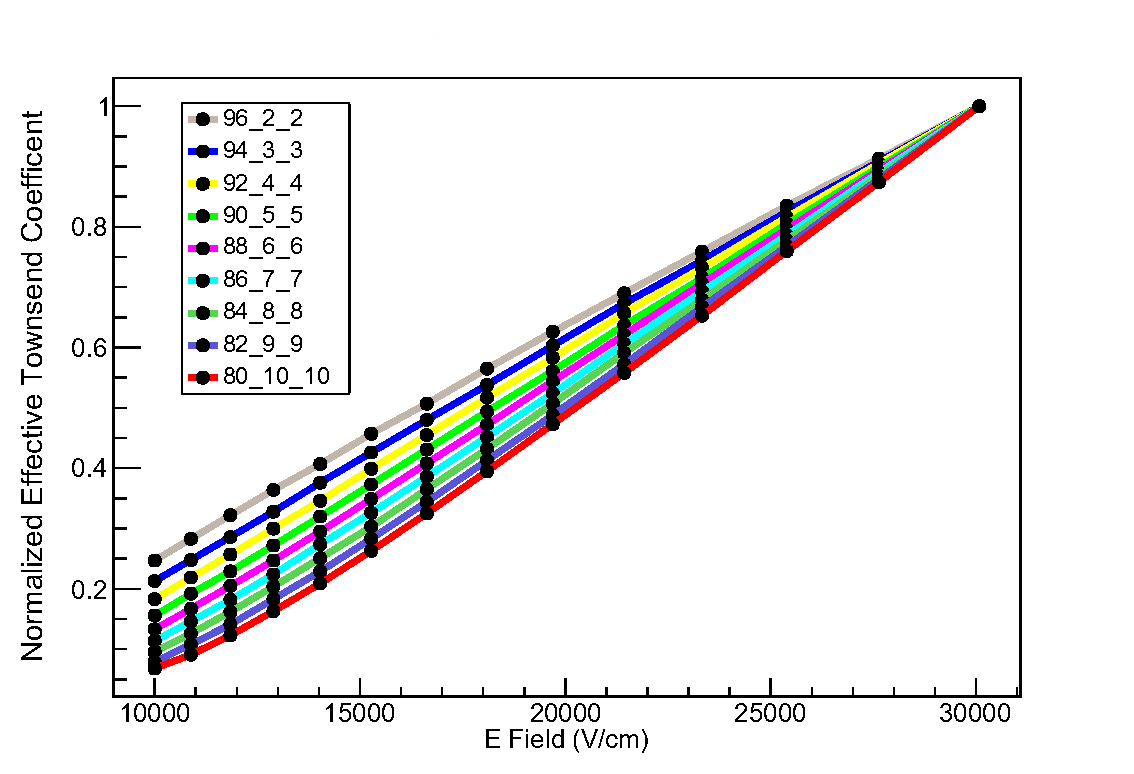}
\caption{Simulation of normalization effective townsend coefficient versus electric field across various gas mixtures.\label{fig:f6}}
\end{figure}

The $ \alpha _{eff} $ for each gas mixture was calculated when the electric field varied from 10 kV/cm to 30 kV/cm. For each gas mixture, the Penning transfer is fixed at 0.7, consistent with the settings used in the prior simulations. Figure \ref{fig:f6} shows the normalization $ \alpha _{eff} $ for various gas mixtures as a function of the electric field, with $ \alpha _{eff} $ set to 1 at 30 kV/cm for each gas mixture. The absolute value of $ \alpha _{eff} $ increases exponentially with the increase of the electric field, though the growth trend varies differently under different gas conditions. The result from $ \alpha _{eff} $ shows that as the electric field fluctuates, a gas mixture with a higher neon percentage reacts less sensitively to the variation. This observation explains the similar variation trends observed in the detector gain.

\subsection{Experimental Results}
\begin{figure}[htbp]
\centering
\includegraphics[width=1.0\textwidth]{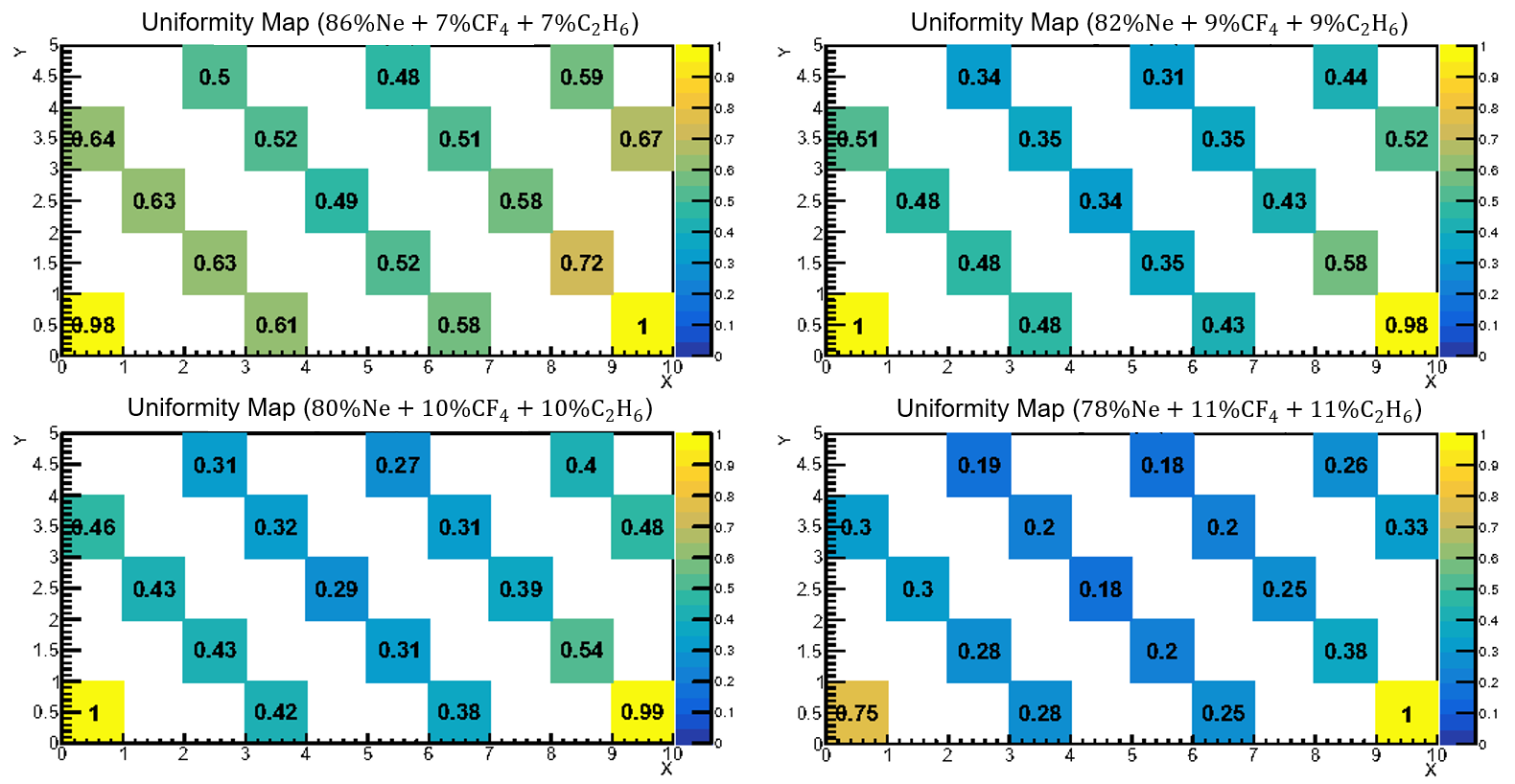}
\caption{Experimental result of gain uniformity map across various gas mixtures.\label{fig:f7}}
\end{figure}

In the test, the detector's uniformity was evaluated by testing a subset of its channels. Several channels from the upper part of the detector were selected to represent the entire detector since measuring each channel of the detector in a single run is time-consuming. During the test, high voltages of 550 V and 300 V are applied to the pre-amp and amplification gap, respectively. Figure \ref{fig:f7} presents the results of testing the uniformity of the detector across varying neon concentrations in different gas mixtures. It displays the gain distribution across various channels for four gas mixtures. Each gain value is normalized to its highest value, which is set to 1 to facilitate comparison. The four gas mixtures—78\% Ne + 11\% $\text{CF}_4$ + 11\% $\text{C}_2\text{H}_6$, 80\% Ne + 10\% $\text{CF}_4$ + 10\% $\text{C}_2\text{H}_6$, 82\% Ne + 9\% $\text{CF}_4$ + 9\% $\text{C}_2\text{H}_6$  and 86\% Ne + 7\% $\text{CF}_4$ + 7\% $\text{C}_2\text{H}_6$—are tested with uniformities of 60\%, 47\%, 41\%, and 24\%, respectively. Gain uniformity is determined by calculating the ratio of the standard deviation to the mean value of each channel. The results indicate that uniformity decreases rapidly as the proportion of neon increases. Specifically, the uniformity improves 2.5 times better when the detector operates with a gas mixture containing 86\% neon compared to 78\%. The improvement in uniformity with higher neon concentration shows the same trend as the simulation results obtained earlier.

\section{Timing Performance}
\begin{figure}[htbp]
\centering
    \begin{minipage}[t]{0.45\textwidth} 
        \centering 
        \includegraphics[width=\textwidth]{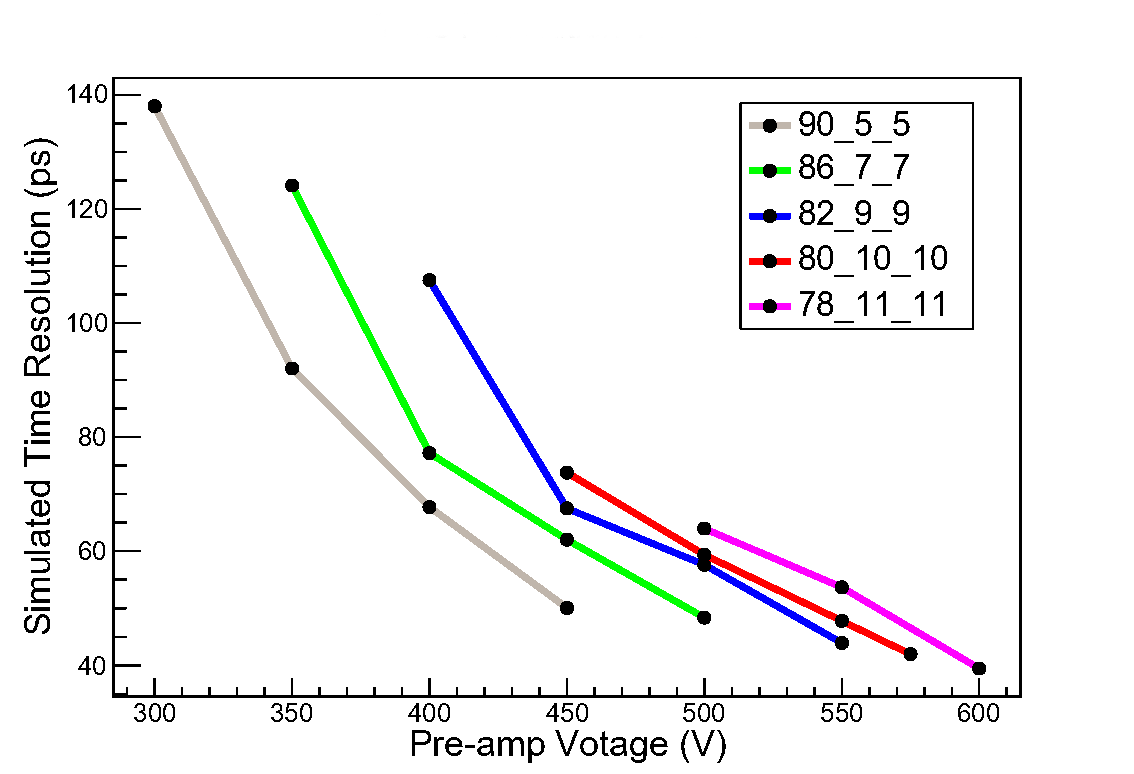} 
        \vspace{2mm} 
        \textbf{(a)} 
    \end{minipage}
    \qquad
    \begin{minipage}[t]{0.45\textwidth}
        \centering 
        \includegraphics[width=\textwidth]{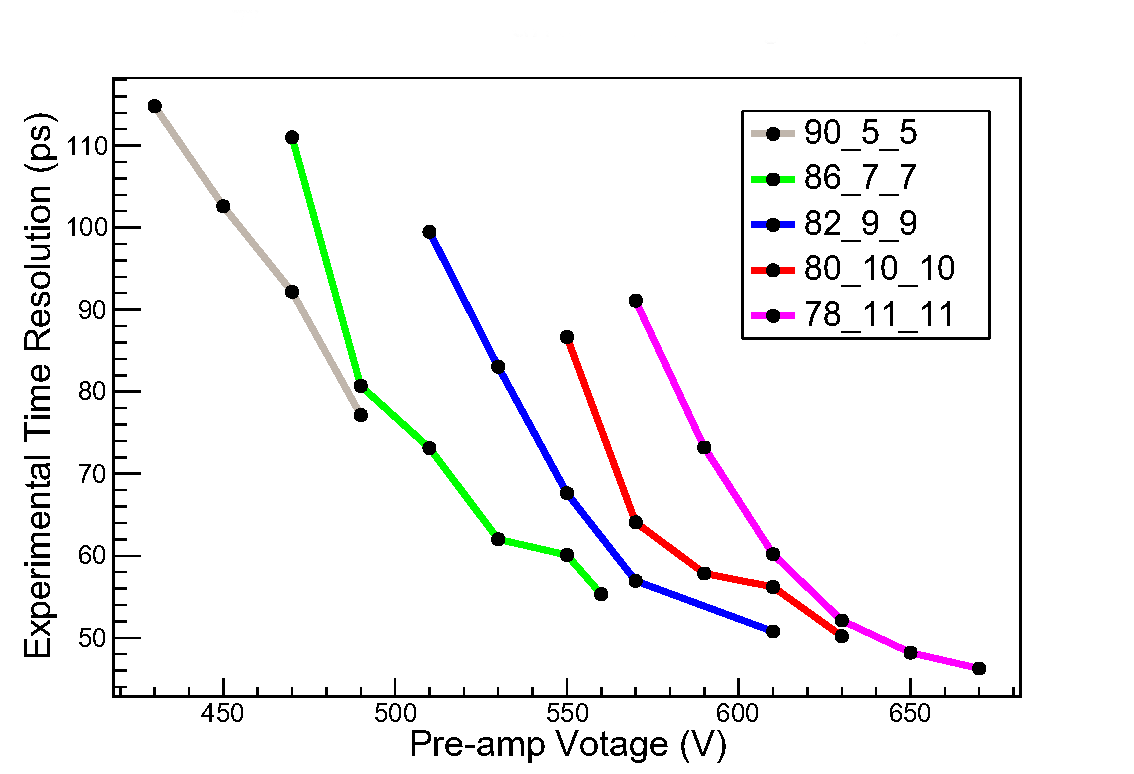} 
        \vspace{2mm} 
        \textbf{(b)} 
    \end{minipage}
\caption{(a) Simulated and (b) Experimental time resolution versus pre-amp voltage across various gas mixtures.\label{fig:f8}}
\end{figure}

The detector's time resolution was evaluated to assess its timing performance, with the simulation and experimental results showing a similar trend. Figure \ref{fig:f8} (a) and (b) depict simulation and experimental time resolution as a function of pre-amp voltage across different gas mixtures, respectively. During both simulation and testing, the high voltage applied to the amplification region is fixed at 300 V, while the voltage on the pre-amp gap varies. In practical conditions, the high voltage applied to the pre-amp gap is limited. Therefore, the last point of each curve in the experimental data represents the highest voltage that can be applied to the pre-amp gap while maintaining detector stability.

The results reveal that at the same pre-amp voltage, a higher concentration of neon leads to improved time resolution. This can be attributed to the timing of the first ionization that occurred in the pre-amplification gap. Previous studies have shown that earlier ionization results in a higher electron drift velocity, which in turn enhances the overall timing performance \cite{p}. With the same applied voltage, a higher proportion of neon leads to earlier first ionization due to the increased ionization probability, thereby improving time resolution. On the contrary, reducing the neon concentration in the gas mixture allows the pre-amplification gap to tolerate higher voltages, which leads to better achievable time resolution at the maximum limit voltage. For all gas mixtures shown, the time resolution decreases as the voltage increases, which is consistent with both experimental and simulation results. In the experiments, the declination in the time resolution slows down at very high voltages, possibly due to performance limitations such as discharge issues in the detector.

\section{Conclusion}
In this work, we investigated the influence of various gas mixtures on the gain uniformity and time resolution of the multi-channel PICOSEC MM detector. A Monte Carlo simulation was employed to evaluate detector performance across several gas mixtures, and experiments were conducted with the multi-channel PICOSEC MM to validate the simulation results. The experimental results showed consistency with the simulations. Both simulation and experimental results indicate that increasing the concentration of neon reduces the detector's sensitivity to variations in the pre-amp gap, leading to improved uniformity. Specifically, increasing the neon concentration from 78\% to 86\% resulted in a 2.5-fold improvement in uniformity. Conversely, we found that better time resolution is more achievable with a lower concentration of neon. It is important to acknowledge that the working gas impacts both the gain uniformity and the timing performance of the detector but in opposite directions regarding performance optimization. Therefore, future applications will require a careful balance of these two aspects based on the detector's operating conditions and performance requirements to identify the optimal working gas. This study offers valuable insights into enhancing the uniformity of large-area PICOSEC MM and provides guidance for selecting an optimal working gas to achieve peak detector performance in future practical applications.

\acknowledgments

This work was supported by the Program of National Natural Science Foundation of China (grant number 11935014, 12125505, 12075238). We acknowledge the pioneering work on developing the PICOSEC detector concept by the PICOSEC collaboration, which provided a foundational framework for this research. We are grateful to the PICOSEC collaboration for their help and support during this research.


\bibliographystyle{JHEP}
 \bibliography{biblio.bib}


%
\end{document}